\documentclass[conference]{IEEEtran}

\usepackage[USenglish]{babel}
\usepackage[utf8]{inputenc}
\usepackage{amstext}
\usepackage{amsmath}
\usepackage{graphicx}
\usepackage{amsfonts}
\usepackage{amssymb}
\usepackage{enumerate}
\usepackage{mathdots}
\usepackage{comment}
\usepackage{setspace}
\usepackage{float}
\usepackage{multicol}
\usepackage{slashbox}
\usepackage{multirow}
\usepackage[table]{xcolor}
\usepackage{hhline}

\DeclareMathOperator{\sign}{sign}
\newcommand{\RE}{\mathbb{R}}

\newtheorem{remark}{Remark}
\newtheorem{assumption}{Assumption}

\begin{document}
\title{Is It Reasonable to Substitute Discontinuous SMC by Continuous HOSMC? \\ {\LARGE Comments to Discussion Paper by V. Utkin}}

\author{\IEEEauthorblockN{Ulises P\'{e}rez Ventura}
\IEEEauthorblockA{Posgrado de Ingenier\'{i}a \\ Facultad de Ingenier\'{i}a, UNAM, M\'{e}xico\\
Email: ventury.sk8@gmail.com}
\and
\IEEEauthorblockN{Leonid Fridman}
\IEEEauthorblockA{Departamento de Ingenier\'{i}a de Control y Rob\'{o}tica \\ Facultad de Ingenier\'{i}a, UNAM, M\'{e}xico\\
Email: lfridman@unam.mx}}

\maketitle

\begin{equation*}
\vspace{-2cm}
\end{equation*}

\begin{abstract}
Professor Utkin in his discussion paper proposed an example showing that the amplitude of chattering caused by the presence of parasitic dynamics in systems governed by First-Order Sliding-Mode Control (FOSMC) is lower than the obtained using Super-Twisting Algorithm (STA). This example served to motivate this research reconsidering the problem of comparison of chattering magnitude in systems governed by FOSMC that produces a discontinuous control signal and by STA that produces a continuous one, using Harmonic Balance (HB) methodology. With this aim the Averaged Power (AP) criteria for chattering measurements is revisited. The STA gains are redesigned to minimize amplitude or AP of oscillations predicted by HB. The comparison of the chattering produced by FOSMC and STA with redesigned gains is analyzed taking into account their amplitudes, frequencies and values of AP allowing to conclude that: (a) for any value of upperbound of disturbance and Actuator Time Constant (ATC) there exist a bounded disturbance for which the amplitude and AP of chattering produced by FOSMC is lower than the caused by STA; (b) if the upperbound of disturbance and upperbound of time-derivative disturbance are given, then for all sufficiently small values of ATC the amplitude of chattering and AP produced by STA will be smaller than the caused by FOSMC; (c) critical values of ATC are predicted by HB for which the parameters, amplitude of chattering and AP, produced by FOSMC and STA are the same. Also the frequency of self-exited oscillations caused by FOSMC is always grater than the produced by STA.   
\end{abstract}

\section{Introduction}
Sliding-Mode Control (SMC) is an efficient technique used for bounded matched uncertainty compensation \cite{Utkin92}. The First-Order Sliding-Mode Control (FOSMC) keeps a desired constraint $\sigma$ of relative degree one, by means of theoretically infinite-frequency switching control. However, infinite-frequency switching control is not feasible due to the presence of parasitic dynamics as actuators and sensors \cite{Fridman01} hysteresis effects \cite{Tsypkin85}, \cite{Boiko09}, and other non-idealities. Hence, the sliding set converges to a real sliding motion with finite (high) frequency, this effect is well-known as chattering effect and it is the main drawback of the sliding-mode control theory. \\

Higher-Order Sliding Mode Control (HOSMC) algorithms were proposed as an attempt to adjust the chattering by substituting (intuitively) discontinuous control inputs by continuous ones \cite{Emelyanov86}, \cite{Levant93}, \cite{Bartolini98}. One of the most efficient algorithms is the Super-Twisting Algorithm (STA) compensating theoretically 

\begin{equation*}
\vspace{-2.1cm}
\end{equation*}

\noindent
exactly matched Lipschitz uncertainties in finite-time \cite{Levant93}, \cite{Levant98}. This idea was very attractive but in the paper \cite{Boiko05} was shown that in systems driven by STA (as well as by any other controller with infinite gain at the origin) the chattering also appears. Moreover, Professor Utkin \cite{Utkin15}, \cite{Utkin16}, \cite{Swikir16}, presented some examples showing that some systems governed by STA exhibit bigger amplitude of chattering that the systems driven by First-Order Sliding Mode Control (FOSMC), when parasitic dynamics affects the SMC/HOSMC closed-loop. \\

In this paper we will try to answer the question: \textit{Is It Expedient to Substitute Discontinuous SMC by Continuous HOSMC}, when parasitic dynamics are presented in the control loop? For that, we analyze the amplitude and frequency of possible self-excited oscillations and its effect on the Average Power (AP). Harmonic Balance (HB) is widely used to estimate the amplitude and frequency of possible oscillations for dynamically perturbed SMC/HOSMC systems \cite{Tsypkin85}, \cite{Boiko05}, where a Describing Function (DF) characterizes the non-linearities effects on the parameters of periodic motions \cite{Gelb68}, \cite{Atherton75}. \\

\noindent 
The contributions of this work are listed below:
\begin{itemize}
\item We confirm the hypothesis of Professor V. Utkin: for any value of the actuator time-constant (ATC) there exist a bounded disturbance for which the amplitude of possible oscillations produced by FOSMC is lower than the obtained applying STA. 
\item Given the upperbound of disturbance and the upperbound of time-derivative disturbance, for a sufficiently small value of ATC the amplitude of chattering and AP produced by STA will be smaller than the caused by FOSMC.
\end{itemize}
\noindent 
With this aim, we propose the following:
\begin{itemize}
\item[1.] Reformulation of A. Levant ``energy like" criterion to compute the AP based on HB methodology.
\item[2.] Selection of STA gains to minimize the amplitude of  possible oscillations.
\item[3.] Selection of STA gains to minimize the AP.\\
\end{itemize}

The structure of the paper is as follows: a motivation example is discussed in section II; section III contains the preliminaries about HB approach for FOSMC and STA; chattering parameters obtained by HB are analyzed in section IV; comparison examples in section V; and section VI presents the conclusions about the obtained results.

\section{Motivation Example}
Consider the disturbed first-order system shown in the Figure 1, the plant can be modeled as
\begin{equation}\label{System}
\dot{x}(t) = \bar{u}(t) + F(t) \,,
\end{equation}
where $x \, \in \, \RE$ is the output and $\bar{u} \, \in \, \RE$ is the control input. The disturbance term $F$ has the form
\begin{equation}\label{Disturbance}
F  =  \alpha \sin (\Omega t) \hspace{4.5mm} \Rightarrow \hspace{4.5mm} \left\lbrace
\begin{array}{l}
|F| \leq \delta = \alpha \\
|\dot{F}| \leq \Delta = \alpha \, \Omega
\end{array} \right.
\end{equation}

\begin{figure}[t]
\begin{center}\label{Bloques}
\vspace{-2mm}
\includegraphics[scale=0.22]{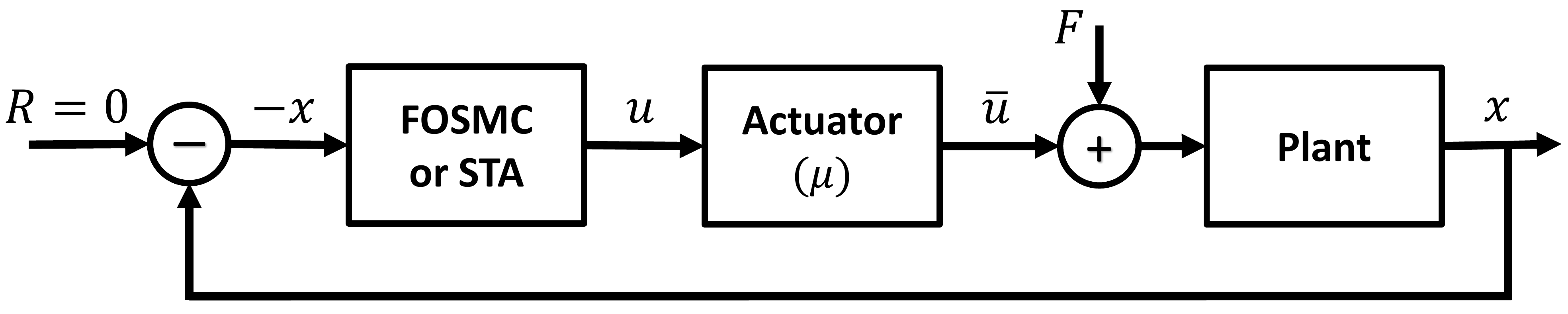}   
\end{center}
\vspace{-4mm}
\caption{Control Scheme.}
\vspace{-4mm}
\end{figure}

\noindent
Let us apply two control laws:

\begin{itemize}
\item Discontinuous FOSMC \cite{Utkin92}: \vspace{2mm}
\end{itemize}
\begin{equation}\label{SGN}
u = - M\sign (x) \,,
\end{equation}
where the control gain is chosen $M = 1.1\delta$ from the upperbound of disturbance (\ref{Disturbance}), ensuring the global finite-time convergence to the first order sliding-mode ($\exists \, t_r: x(t)=0$, $\forall \, t \geq t_r$) when the actuator dynamics is fast enough. 

\begin{itemize}
\item Continuous STA \cite{Levant93}, \cite{Levant98}: \vspace{2mm}
\end{itemize}
\begin{equation}\label{STA}
\begin{array}{ccl}
u & = & - k_1 |x|^{1/2}\sign (x) + v \,, \\
\dot{v} & = & -k_2\sign (x) \,,
\end{array}
\end{equation}
where the control gains are chosen $k_1 = 1.5\sqrt{\Delta}$, $k_2 = 1.1\Delta$ from the upperbound of time-derivative disturbance (\ref{Disturbance}), ensuring the global finite-time convergence to the second order sliding-mode ($\exists \, t_r: x(t)=\dot{x}(t)=0$, $\forall \, t \geq t_r$) when the actuator dynamics is fast enough.

\vspace{1mm}
\begin{remark}
\textit{FOSMC can reject bounded disturbances but STA can compensate Lipschitz disturbances (not necessarily bounded). In order to compare both algorithms we consider a bounded and Lipschitz disturbance (\ref{Disturbance}).}
\end{remark}
\vspace{1mm}

Following \cite{Utkin15}, we consider the same actuator model as Professor V. Utkin which consists of a 2$^{nd}$-order linear system
\begin{equation}\label{Actuator}
\begin{array}{rcl}
\dot{z}(t) & = & \begin{bmatrix} 0 & 1 \\ -\frac{1}{\mu^2} & -\frac{2}{\mu} \end{bmatrix}z(t) + \begin{bmatrix} 0 \\ \frac{1}{\mu^2} \end{bmatrix}u(t) \,, \vspace{2mm}\\
\bar{u}(t) & = & \begin{bmatrix} 1 & 0 \end{bmatrix}  z(t) \,,
\end{array}
\end{equation}
where $\mu>0$ is the actuator time-constant (ATC). Thus, the effects of parasitic dynamics can be parameterized through ATC. Let us notice that FOSMC and STA were designed for the system (\ref{System}) with relative degree one but the presence of actuator (\ref{Actuator}) increases the relative degree of the system (\ref{System}). \\

Table \ref{Accuracy} presents the simulation results of the system conformed by the plant (\ref{System}) and the actuator dynamics (\ref{Actuator}) in closed loop with FOSMC (\ref{SGN}) and STA (\ref{STA}). Chattering magnitude of the output $x$ is compared taking into account several values of ATC $\mu$ and disturbance frequency $\Omega$ (fixing $\alpha = 1$). It can be seen that the amplitude of chattering has the same order for $\Omega = 1$ when $\mu = 10^{-1}$, and it is lower for STA when $\mu = 10^{-2}$ and $\mu = 10^{-3}$. For $\Omega = 10$ the results change, when $\mu = 10^{-1}$ the amplitude of oscillations for FOSMC is lower than for STA, when $\mu = 10^{-2}$ they have the same order, but when $\mu = 10^{-3}$ the amplitude of chattering in the system governed by STA is lower that ones in the system with FOSMC. The third column of Table \ref{Accuracy} shows that the amplitude of oscillations for FOSMC and STA have the same order only for $\mu=10^{-3}$ and for bigger values of ATC the amplitude generated by FOSMC is lower than the produced by STA. \\

\subsubsection*{Conclusions}
\vspace{-2mm}
\begin{itemize}
\item Table \ref{Accuracy} confirms the hypothesis of Professor V. Utkin: for any value of ATC there exist a bounded disturbance for which the amplitude of possible oscillations produced by FOSMC is lower than the obtained applying STA.
\item It should exists a value of ATC for which the amplitude of chattering produced by FOSMC and STA are the same.
\item For any bounded disturbance, the amplitude of possible oscillations produced by STA may be less than the obtained using FOSMC if the actuator dynamics is fast enough ($\mu \to 0$).
\end{itemize}

\begin{table}[]
\centering
\scalebox{1.08}{
\begin{tabular}{|c|c||c|c|c|}
\hline
% Signo
\multicolumn{2}{|c||}{\bf \backslashbox{{\tiny Control}}{{\tiny $\Omega$}}} & {\tiny \textbf{1}} & {\tiny \textbf{10}} & {\tiny \textbf{100}} \\ \hline \hline
\multicolumn{5}{|c|}{{\tiny Discontinuous Control}} \\ \hline \hline
% Signo
\multirow{3}{*}{{\tiny \textbf{FOSMC}}} & {\tiny $\mathbf{\mu = 10^{-1}}$} & \begin{tiny} \cellcolor{blue!30} 1.366$\times 10^{-1}$ \end{tiny} & \begin{tiny} 1.692$\times 10^{-1}$  \end{tiny} & \begin{tiny} 0.934$\times 10^{-1}$  \end{tiny} \\ \hhline{*{1}{~}*{4}{|-}} & {\tiny $\mathbf{\mu = 10^{-2}}$} & \begin{tiny} 1.092$\times 10^{-2}$ \end{tiny} & \begin{tiny} \cellcolor{blue!20} 1.361$\times 10^{-2}$  \end{tiny} & \begin{tiny} 1.692$\times 10^{-2}$  \end{tiny} \\ \hhline{*{1}{~}*{4}{|-}} & {\tiny $\mathbf{\mu = 10^{-3}}$} & \begin{tiny} 1.064$\times 10^{-3}$ \end{tiny} & \begin{tiny} 1.096$\times 10^{-3}$ \end{tiny} & \begin{tiny} \cellcolor{blue!10} 1.362$\times 10^{-3}$ \end{tiny} \\ \hline \hline
\multicolumn{5}{|c|}{{\tiny Continuous Control}} \\ \hline \hline
% STA
\multirow{3}{*}{{\tiny \textbf{STA}}} & {\tiny $\mathbf{\mu = 10^{-1}}$} & \begin{tiny} \cellcolor{blue!30} 1.243$\times 10^{-1}$ \end{tiny} & \begin{tiny} 8.663$\times 10^{-1}$  \end{tiny} & \begin{tiny} 6.4041  \end{tiny} \\ \hhline{*{1}{~}*{4}{|-}} & {\tiny $\mathbf{\mu = 10^{-2}}$} & \begin{tiny} 9.431$\times 10^{-4}$ \end{tiny} & \begin{tiny} \cellcolor{blue!20} 1.302$\times 10^{-2}$ \end{tiny} & \begin{tiny} 8.694$\times 10^{-2}$ \end{tiny} \\ \hhline{*{1}{~}*{4}{|-}} & {\tiny $\mathbf{\mu = 10^{-3}}$} & \begin{tiny} 8.915$\times 10^{-6}$ \end{tiny} & \begin{tiny} 9.445$\times 10^{-5}$ \end{tiny} & \begin{tiny} \cellcolor{blue!10} 1.343$\times 10^{-3}$ \end{tiny} \\ \hline
\end{tabular}}
\vspace{2mm}
\caption{Sliding-Mode Amplitude Accuracy Increasing the Frequency of Disturbance.}\label{Accuracy}
\vspace{-8mm}
\end{table}

\section{Chattering Analysis of FOSMC and STA Using Harmonic Balance}
Taking into account the control scheme shown in Figure 1, FOSMC (\ref{SGN}) and STA (\ref{STA}) are analyzed in frequency domain using the HB methodology to understand how the parasitic dynamics (\ref{Actuator}) may degrade the accuracy, when the control gains are selected to reject the matched disturbance $F$. \\

Consider the nominal case ($F=0$), the dynamically perturbed system (\ref{Actuator})-(\ref{System}) has the transfer function 
\begin{equation}\label{System_DinPert}
W(s) = G_a(s)G(s) =\dfrac{1}{s(\mu s+1)^2} \,, 
\end{equation}
whose relative degree is $r=3$. \vspace{2mm}

\begin{assumption}
\textit{Due the parasitic dynamics (\ref{Actuator}), the output of the system converges to a periodic solution \cite{Boiko05}, \cite{Utkin15}, which can be approximated by its first-harmonic,
\begin{equation}\label{Limit_Cycle}
\begin{array}{lcl}
x(t) = A \sin(\omega t) \,, \\
\dot{x}(t) = A \omega \cos(\omega t)\,. 
\end{array}
\end{equation}
where $A$ and $\omega$ are the chattering parameters, amplitude and frequency, respectively.} 
\end{assumption}

\subsection{Amplitude and Frequency Estimation}
Let us apply the DF method \cite{Gelb68}, \cite{Atherton75} to predict periodic oscillations. Parameters of a possible limit cycle may be found as an intersection point of the actuator-plant dynamics $W(s)$ Nyquist plot and the negative reciprocal DF $-N^{-1}(A,\omega)$ of the SMC/HOSMC algorithm, which corresponds to the Harmonic Balance equation (HBE)  
\begin{equation}\label{HBE}
N(A,\omega) W(j\omega) = -1 \,,
\end{equation}  
whose solution is an estimate of chattering parameters: amplitude $A$ and frequency $\omega$. 

\subsection{Averaged Power Criteria}
In the paper by A. Levant \cite{Levant10} an ``energy like" criteria for chattering measurement are presented
\begin{equation}
\text{E} = \left( \int_{0}^{T} \dot{x}^p(\tau) d\tau \right)^{1/p} \,,
\end{equation}
inspired by $L_p$ norm. Unfortunately, it is only a qualitative criterion to understand the chattering effects because it has no physical sense and require information of $\dot{x}$ (knowledge of disturbance) to compute it. \\

HB approach allows to compute the Averaged Power (AP) of the steady-state behavior of the system,  
\begin{equation}\label{Averaged_P}
\text{P} = \dfrac{\omega}{2\pi} \int_{0}^{\frac{2\pi}{\omega}} \Bigl( A\omega\cos(\omega t) \Bigl)^2  dt = \dfrac{A^2\omega^2}{2} \,,
\end{equation}
due to the output $x$ and its time-derivative $\dot{x}$ are assumed of the form (\ref{Limit_Cycle}).

\subsection{HB Analysis of First Order Sliding Mode Control}
The describing function of the non-linearity (\ref{SGN}) has the form \cite{Gelb68}
\begin{equation}\label{DF_SGN}
N(A) = \dfrac{4 M}{\pi A} \,.
\end{equation}
The HBE (\ref{HBE}) can be separated as real and imaginary parts
\begin{equation*}
\begin{array}{ccl}
\dfrac{4 M}{\pi A} & = & 2 \, \mu \, \omega^2 \,, \vspace{2mm} \\ 
0 & = & \omega(\mu^2\omega^2-1) \,,
\end{array}
\end{equation*}
whose analytic solution is \cite{Swikir16}
\begin{eqnarray}
A & = & \frac{2 M \mu}{\pi}, \label{A_SGN}\\
\omega & = & \frac{1}{\mu}. \label{w_SGN}
\end{eqnarray}
Substituting the limit cycle parameters (\ref{A_SGN}), (\ref{w_SGN}) in the expression (\ref{Averaged_P}), the AP has the form
\begin{equation}\label{P_SGN}
\text{P} = \dfrac{2M^2}{\pi^2} \,.
\end{equation}

\begin{figure}[t]
\begin{center}\label{Min_A_P}
\includegraphics[scale=0.59]{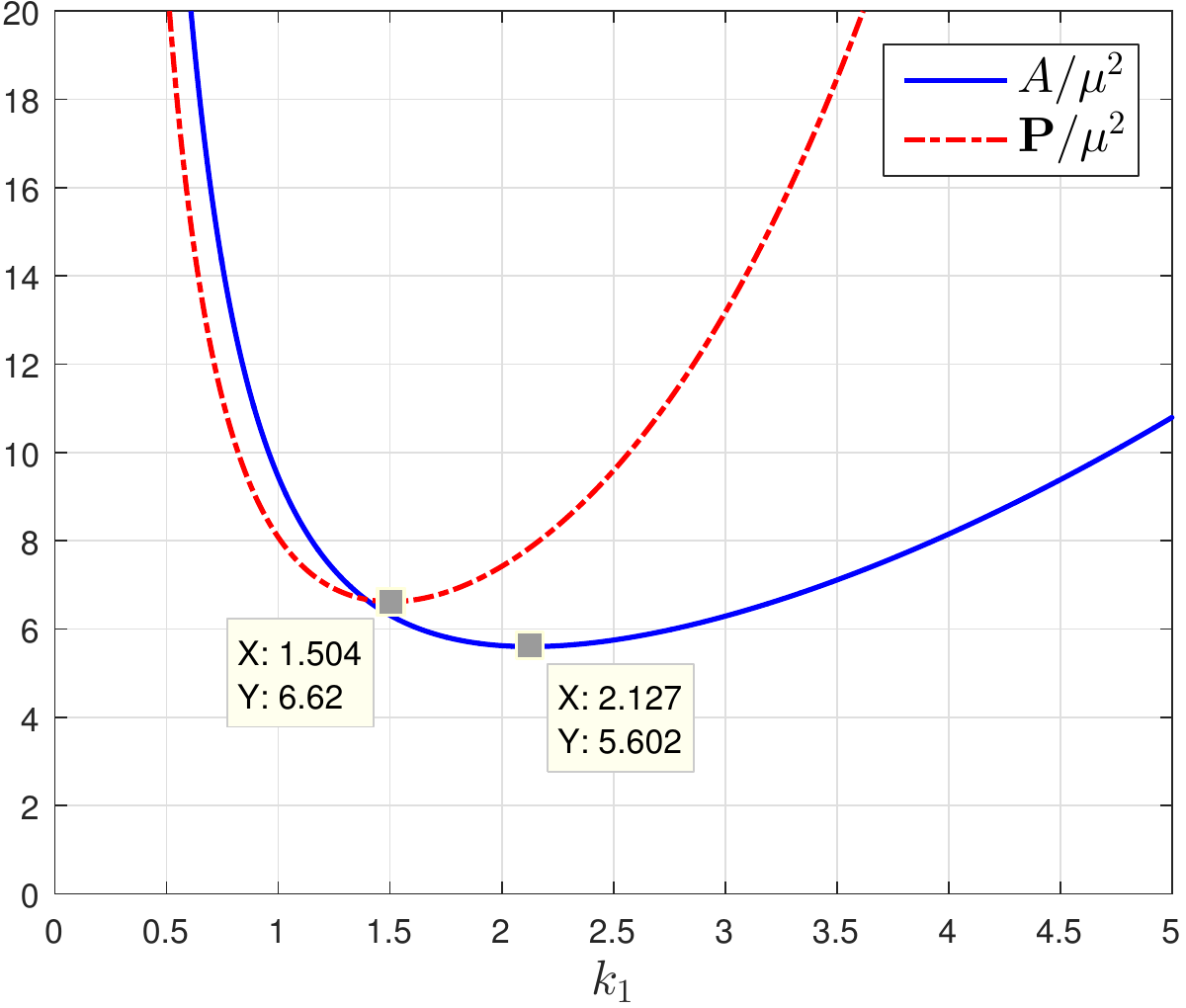}   
\end{center}
\vspace{-4mm}
\caption{Critical points for the normalization of amplitude (\ref{Normal_A}) and AP (\ref{Normal_P}), respectively.}
\vspace{-4mm}
\end{figure}

\subsection{HB Analysis of Super-Twisting Algorithm}
The describing function of the non-linearity (\ref{STA}) has the form \cite{Boiko05}
\begin{equation}\label{DF_STA}
N(A,\omega) = \dfrac{1.1128 k_1}{A^{1/2}} - j \dfrac{4 k_2}{\pi A \omega} \,.
\end{equation}
Once again, the HBE (\ref{HBE}) can be separated as real and imaginary parts
\begin{equation*}
\begin{array}{ccl}
\dfrac{1.1128k_1}{A^{1/2}} & = & 2 \, \mu \, \omega^2 \,, \vspace{2mm} \\ 
\dfrac{4k_2}{\pi A \omega} & = & \omega(1-\mu^2\omega^2) \,,
\end{array}
\end{equation*}
whose analytic solution is \cite{Ventura16}
\begin{eqnarray}
A & = & \mu^2 \mathbb{K}_{_A}, \label{A_STA} \\
\omega & = & \frac{\mathbb{K}_\omega}{\mu}, \label{w_STA}
\end{eqnarray} 
where
\begin{equation*}
\begin{array}{rcl}
\mathbb{K}_A & = & \left( \frac{1}{2}\cdot\frac{(1.1128k_1)^2 + \frac{16}{\pi}k_2}{1.1128k_1}  \right)^2 = \left( \frac{1.1128k_1}{2\mathbb{K}_\omega^2} \right)^2 \,, \vspace{2mm} \\
\mathbb{K}_\omega & = & \left( \frac{(1.1128k_1)^2}{(1.1128k_1)^2 + \frac{16}{\pi}k_2} \right) ^{1/2} \,. 
\end{array}
\end{equation*}
Substituting the limit cycle parameters (\ref{A_STA}), (\ref{w_STA}) in the expression (\ref{Averaged_P}), the AP has the form
\begin{equation}\label{P_STA}
\text{P} = \dfrac{\mu^2}{32} \cdot \dfrac{ \left( \, (1.1128k_1)^2 + \frac{16}{\pi}k_2 \, \right)^{3} }{(1.1128k_1)^2} \,.
\end{equation}

\vspace{4mm}

\subsection{Design of STA Gains to Minimize the Amplitude of Possible Oscillations}
The expression (\ref{A_STA}) allows to obtain the values of STA gains, $k_1$ and $k_2$, such that the amplitude is minimized. Consider the following normalization 
\begin{equation}\label{Normal_A}
\dfrac{A}{\mu^2} = \left( \frac{1}{2}\cdot\frac{(1.1128k_1)^2 + \frac{16}{\pi}k_2}{1.1128k_1}  \right)^2 \,,
\end{equation}
it is possible to compute the value of the gain $k_1$ which minimizes the amplitude (\ref{A_STA}) for each $k_2 > \Delta$,
\begin{equation}\label{STA_k1_min}
k_1 = \left( \dfrac{16k_2}{\pi(1.1128)^2} \right)^{1/2} = 2.028\sqrt{k_2} \,.
\end{equation}

\vspace{2mm}
\begin{remark}
\textit{The selection of STA gains that minimize the amplitude of possible oscillations based on HB approach is
\begin{equation}\label{STA_A_min}
k_1 = 2.127\sqrt{\Delta} \,, \hspace{4mm} k_2 = 1.1 \Delta \,.
\end{equation}
For these STA gains, the chattering parameters become}
\begin{equation*}
A = 5.6023\Delta\mu^2 \,, \hspace{4mm} \omega = \frac{1}{\mu\sqrt{2}} \,, \hspace{4mm} \text{P} = 7.8464\Delta^2\mu^2 \,. 
\end{equation*}
\end{remark}
\vspace{1mm}

Figure 2 shows the critical value of the amplitude normalization (\ref{Normal_A}) for the STA gains (\ref{STA_A_min}) with $\Delta = 1$. Table \ref{A_min_Simulation} contains the simulation results for $\Delta = 10$ and $\mu = 10^{-2}$, the STA gains are selected to compare the amplitude of oscillations and confirms the selection criterion (\ref{STA_A_min}). 

\begin{table}[t]
\centering
\scalebox{1.08}{
\begin{tabular}{|c|c||c|c|c|}
\hline
% Signo
\multicolumn{2}{|c||}{\bf \backslashbox{{\tiny Parameters}}{{\tiny $k_1$}}} & {\tiny $1.5\sqrt{\Delta}$} & {\tiny $2.127\sqrt{\Delta}$} & {\tiny $2.5\sqrt{\Delta}$} \\ \hline \hline
% HB
\multirow{3}{*}{{\tiny \textbf{\begin{tabular}[c]{@{}c@{}}Harmonic\\ Balance \end{tabular}}}} & {\tiny $\mathbf{A}$} & \begin{tiny} 6.314$\times 10^{-3}$ \end{tiny} & \begin{tiny} \cellcolor{blue!15} 5.602$\times 10^{-3}$  \end{tiny} & \begin{tiny} 5.750$\times 10^{-3}$  \end{tiny} \\ \hhline{*{1}{~}*{4}{|-}} & {\tiny $\mathbf{\omega}$} & \begin{tiny} 57.632 \end{tiny} & \begin{tiny} \cellcolor{blue!15} 70.711 \end{tiny} & \begin{tiny} 76.164 \end{tiny} \\ \hhline{*{1}{~}*{4}{|-}} & {\tiny $\text{P}$} & \begin{tiny} 6.620$\times 10^{-2}$ \end{tiny} & \begin{tiny} \cellcolor{blue!15} 7.846$\times 10^{-2}$ \end{tiny} & \begin{tiny} 9.589$\times 10^{-2}$ \end{tiny} \\ \hline \hline
% SIM
\multirow{3}{*}{{\tiny \textbf{Simulations}}} & {\tiny $\mathbf{A}$} & \begin{tiny} 6.395$\times 10^{-3}$ \end{tiny} & \begin{tiny} \cellcolor{blue!15} 5.653$\times 10^{-3}$  \end{tiny} & \begin{tiny} 5.799$\times 10^{-3}$  \end{tiny} \\ \hhline{*{1}{~}*{4}{|-}} & {\tiny $\mathbf{\omega}$} & \begin{tiny} 57.099 \end{tiny} & \begin{tiny} \cellcolor{blue!15} 70.299 \end{tiny} & \begin{tiny} 75.781 \end{tiny} \\ \hhline{*{1}{~}*{4}{|-}} & {\tiny $\text{P}$} & \begin{tiny} 6.786$\times 10^{-2}$ \end{tiny} & \begin{tiny} \cellcolor{blue!15} 8.009$\times 10^{-2}$ \end{tiny} & \begin{tiny} 9.784$\times 10^{-2}$ \end{tiny} \\ \hline 
\end{tabular}}
\vspace{2mm}
\caption{Simulations of Minimum Amplitude STA Gains (\ref{STA_A_min}) fixing $k_2 = 1.1\Delta$ and $\Delta = 10$.}\label{A_min_Simulation}
\vspace{-6mm}
\end{table}

\subsection{Design of STA Gains to Minimize the Averaged Power}
The expression (\ref{P_STA}) allows to obtain the values of STA gains, $k_1$ and $k_2$, such that the AP is minimized. Consider the following normalization
\begin{equation}\label{Normal_P}
\dfrac{\text{P}}{\mu^2} = \dfrac{1}{32} \cdot \dfrac{ \left( \, (1.1128k_1)^2 + \frac{16}{\pi}k_2 \, \right)^{3} }{(1.1128k_1)^2} \,,
\end{equation}
it is possible to compute the value of the gain $k_1$ which minimizes the AP for each $k_2 > \Delta$,
\begin{equation}\label{STA_k1_min_U}
k_1 = \left( \dfrac{8k_2}{\pi(1.1128)^2} \right)^{1/2} = 1.434 \sqrt{k_2} \,.
\end{equation}

\begin{remark}
\textit{The selection of STA gains that minimize the AP based on HB approach is
\begin{equation}\label{STA_P_min}
k_1 = 1.504\sqrt{\Delta} \,, \hspace{4mm} k_2 = 1.1 \Delta \,.
\end{equation}
For these STA gains, the parameters of periodic motion become}
\begin{equation*}
A = 6.3025\Delta\mu^2 \,, \hspace{4mm} \omega = \frac{1}{\mu\sqrt{3}} \,, \hspace{4mm} \text{P} = 6.6203\Delta^2\mu^2 \,.
\end{equation*}
\end{remark}
\vspace{1mm}

Figure 2 shows the critical value of the AP normalization (\ref{Normal_P}) with $\Delta = 1$. Table \ref{P_min_Simulation} contains the simulation results for $\Delta = 10$ and $\mu = 10^{-2}$, the STA gains are selected to compare the AP and confirms the selection criterion (\ref{STA_P_min}). 

\vspace{2mm}
\begin{remark}
\textit{Sufficient conditions to guarantee finite-time stability are presented in \cite{Moreno09}, where the STA gains have to satisfy }
\begin{equation}
k_1 > 1.414 \sqrt{k_2} \,, \hspace{8mm} k_2 > \Delta \,, 
\end{equation}
\textit{where $\Delta$ is the upperbound of the time-derivative disturbance (\ref{Disturbance}). STA proposed gains in the expressions (\ref{STA_A_min}) and (\ref{STA_P_min}) ensure finite-time stability when the actuator dynamics is fast enough.}
\end{remark}
\vspace{2mm}

\begin{table}[t]
\centering
\scalebox{1.08}{
\begin{tabular}{|c|c||c|c|c|}
\hline
\multicolumn{2}{|c||}{\bf \backslashbox{{\tiny Parameters}}{{\tiny $k_1$}}} & {\tiny $\sqrt{\Delta}$} & {\tiny $1.504\sqrt{\Delta}$} & {\tiny $2\sqrt{\Delta}$} \\ \hline \hline
% HB
\multirow{3}{*}{{\tiny \textbf{\begin{tabular}[c]{@{}c@{}}Harmonic\\ Balance \end{tabular}}}} & {\tiny $\mathbf{A}$} & \begin{tiny} 9.447$\times 10^{-3}$ \end{tiny} & \begin{tiny} \cellcolor{blue!15} 6.302$\times 10^{-3}$  \end{tiny} & \begin{tiny} 5.623$\times 10^{-3}$  \end{tiny} \\ \hhline{*{1}{~}*{4}{|-}} & {\tiny $\mathbf{\omega}$} & \begin{tiny} 42.547 \end{tiny} & \begin{tiny} \cellcolor{blue!15} 57.735 \end{tiny} & \begin{tiny} 68.502 \end{tiny} \\ \hhline{*{1}{~}*{4}{|-}} & {\tiny $\text{P}$} & \begin{tiny} 8.078$\times 10^{-2}$ \end{tiny} & \begin{tiny} \cellcolor{blue!15} 6.620$\times 10^{-2}$ \end{tiny} & \begin{tiny} 7.420$\times 10^{-2}$ \end{tiny} \\ \hline \hline
% SIM
\multirow{3}{*}{{\tiny \textbf{Simulations}}} & {\tiny $\mathbf{A}$} & \begin{tiny} 9.688$\times 10^{-3}$ \end{tiny} & \begin{tiny} \cellcolor{blue!15} 6.383$\times 10^{-3}$  \end{tiny} & \begin{tiny} 5.676$\times 10^{-3}$  \end{tiny} \\ \hhline{*{1}{~}*{4}{|-}} & {\tiny $\mathbf{\omega}$} & \begin{tiny} 41.789 \end{tiny} & \begin{tiny} \cellcolor{blue!15} 57.203 \end{tiny} & \begin{tiny} 68.076 \end{tiny} \\ \hhline{*{1}{~}*{4}{|-}} & {\tiny $\text{P}$} & \begin{tiny} 8.416$\times 10^{-2}$ \end{tiny} & \begin{tiny} \cellcolor{blue!15} 6.781$\times 10^{-2}$ \end{tiny} & \begin{tiny} 7.573$\times 10^{-2}$ \end{tiny} \\ \hline
\end{tabular}}
\vspace{2mm}
\caption{Simulations of Minimum Averaged Power STA Gains (\ref{STA_P_min}) fixing $k_2 = 1.1\Delta$ and $\Delta=10$.}\label{P_min_Simulation}
\vspace{-6mm}
\end{table}

\section{Analysis of Chattering Parameters Estimated by Harmonic Balance}

\begin{figure}[t]
\begin{center}\label{Total_FOS_min_A}
\includegraphics[scale=0.56]{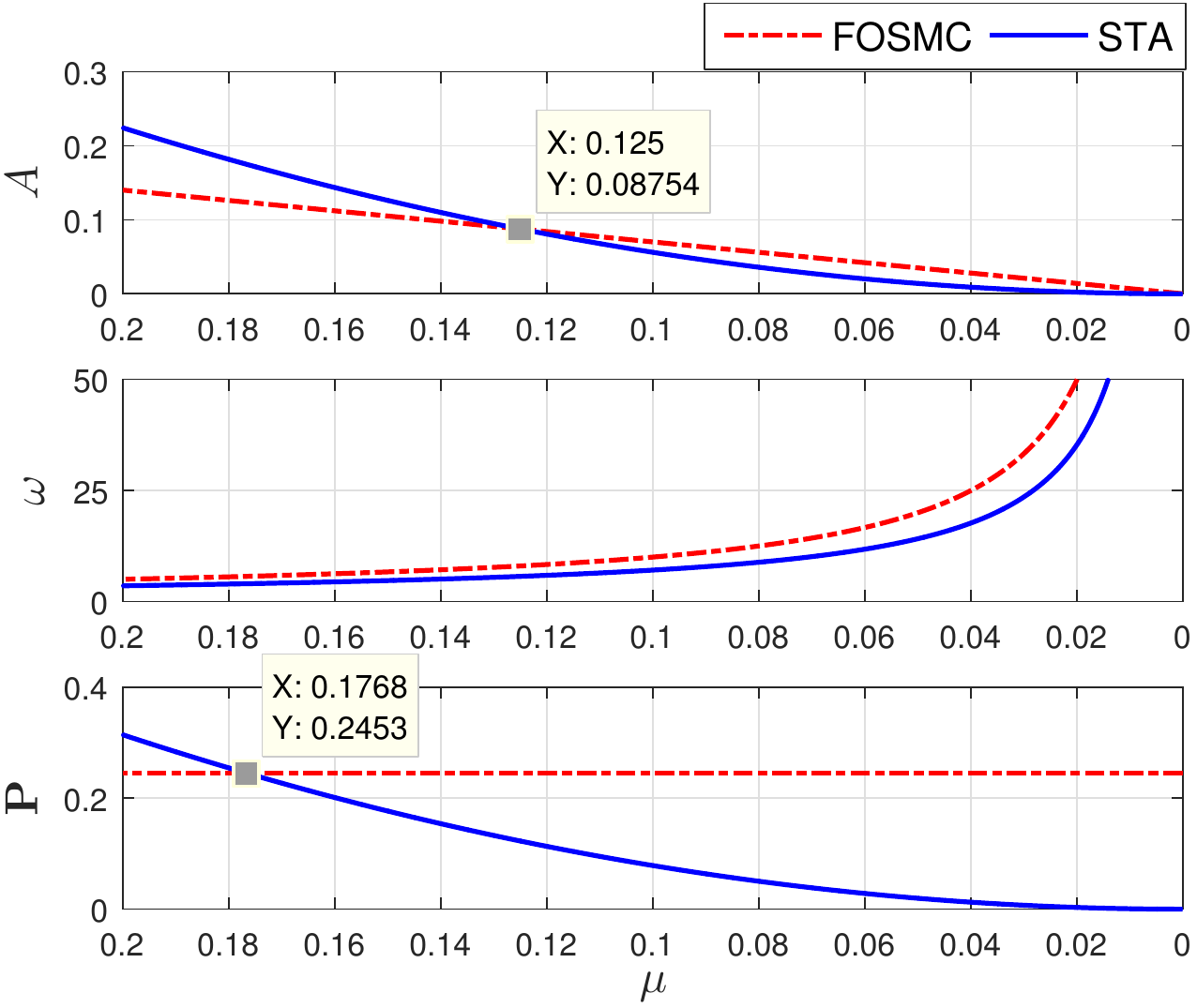}   
\end{center}
\vspace{-4mm}
\caption{Chattering parameters for $\delta = \Delta = 1$ choosing the STA gains (\ref{STA_A_min}).}
\vspace{-4mm}
\end{figure}

\subsection{Amplitude Analysis}
In order to analyze the chattering parameters with respect to the ATC, consider $\delta = \Delta = 1$ and the control gain $M = 1.1\delta$ for FOSMC and the selection proposed in (\ref{STA_A_min}) for STA. Figure 3 shows the chattering parameters for several values of ATC, there is a value of the ATC such that the amplitude of the output is the same  \cite{Ventura16}, despite the use of FOSMC (\ref{SGN}) or STA (\ref{STA}) on the dynamically perturbed system (\ref{System_DinPert}),
\begin{equation}\label{Same_Amplitud}
\mu^\ast = \dfrac{ 8 M (1.1128k_1)^2  }{ \pi \left( \, (1.1128k_1)^2  + \frac{16}{\pi} k_2 \, \right)^2 } \,.
\end{equation}
for this example, the value of ATC to have the same amplitude is
\begin{equation}
\mu^\ast=0.125\dfrac{\delta}{\Delta} \,.
\end{equation} 

\begin{remark}
\textit{Figure 3 confirms that to substitute FOSMC by STA, we should consider that the amplitude of oscillations may be greater(lower) when the ATC,} 
\begin{equation}
\begin{array}{lcl}
\mu > \mu^\ast & \Rightarrow & A_{\text{FOSMC}} < A_{\text{STA}} \,, \\
\mu < \mu^\ast & \Rightarrow & A_{\text{FOSMC}} > A_{\text{STA}} \,. 
\end{array}
\end{equation}
\end{remark}
\vspace{1mm}

\subsection{Frequency Analysis}

\begin{figure}[t]
\begin{center}\label{Nyquist_1}
\includegraphics[scale=0.58]{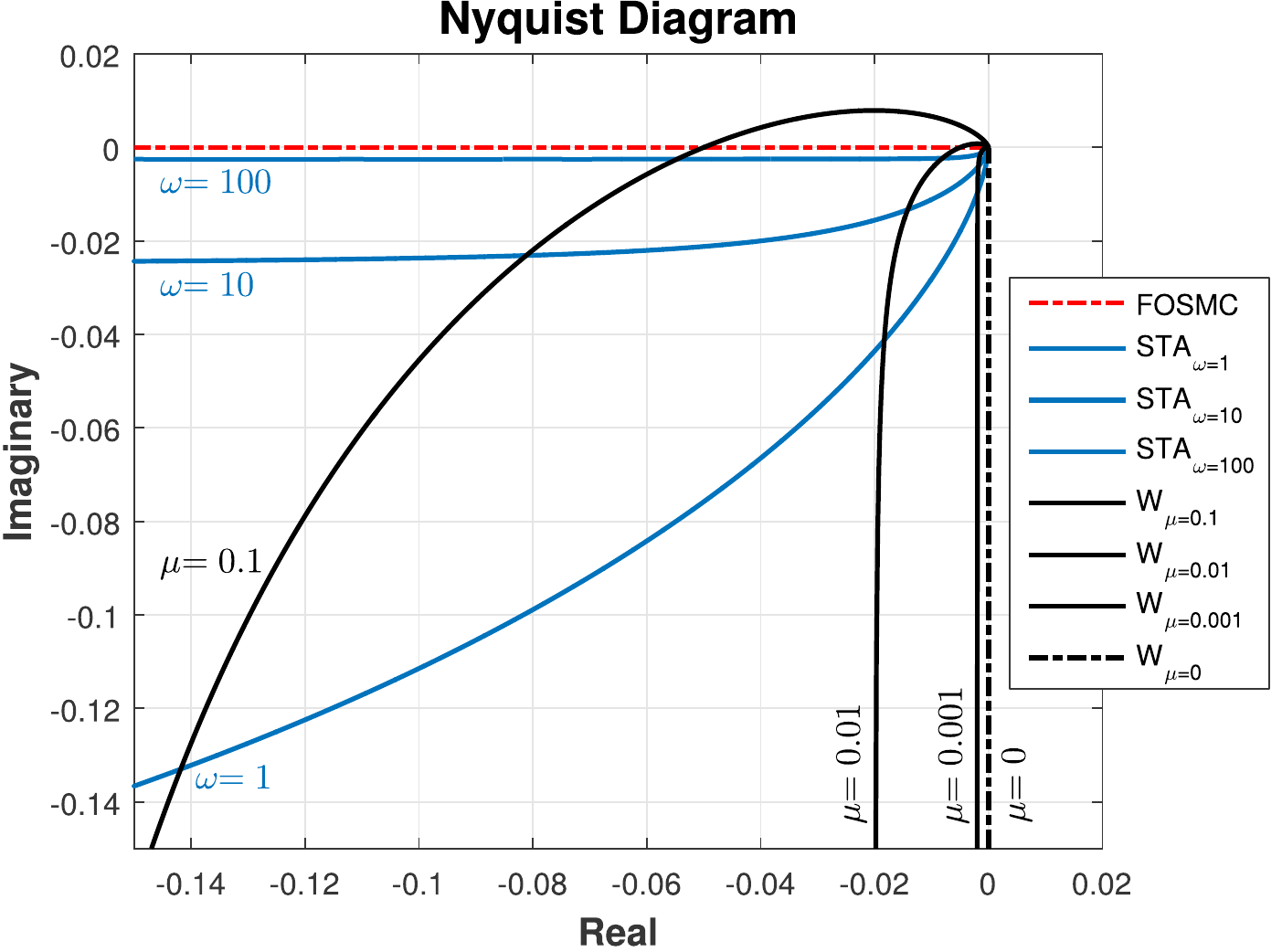}   
\end{center}
\vspace{-5mm}
\caption{Graphical Solution of HB (\ref{HBE}).}
\vspace{-4mm}
\end{figure}

The order of high-frequency oscillations is O$(1/\mu)$ when it is applied FOSMC (\ref{SGN}) or STA (\ref{STA}). However, the frequency is always lower for the STA than the obtained using FOSMC, as it is shown in the graphical solution of the HB equation (\ref{HBE}) of Figure 4.

\subsection{Averaged Power Analysis}
In order to analyze the chattering parameters with respect to the ATC, consider $\delta = \Delta = 1$ and the control gain $M = 1.1\delta$ for FOSMC and the selection proposed in (\ref{STA_P_min}) for STA. Figure 5 shows the chattering parameters for several values of ATC, there is a value of ATC such that the AP is the same despite the use of FOSMC (\ref{SGN}) or STA (\ref{STA}) on the dynamically perturbed system (\ref{System_DinPert}),
\begin{equation}\label{Same_Average_P}
\mu^\star = \dfrac{8M(1.1128k_1)}{\pi \left( \, (1.1128k_1)^2 + \frac{16}{\pi}k_2 \, \right)^{3/2}} \,.
\end{equation} 
for this example, the value of ATC to have the same AP is
\begin{equation}
\mu^\star = 0.1924\dfrac{\delta}{\Delta} \,.
\end{equation} 

\begin{figure}[t]
\begin{center}\label{Total_FOS_min_P}
\includegraphics[scale=0.56]{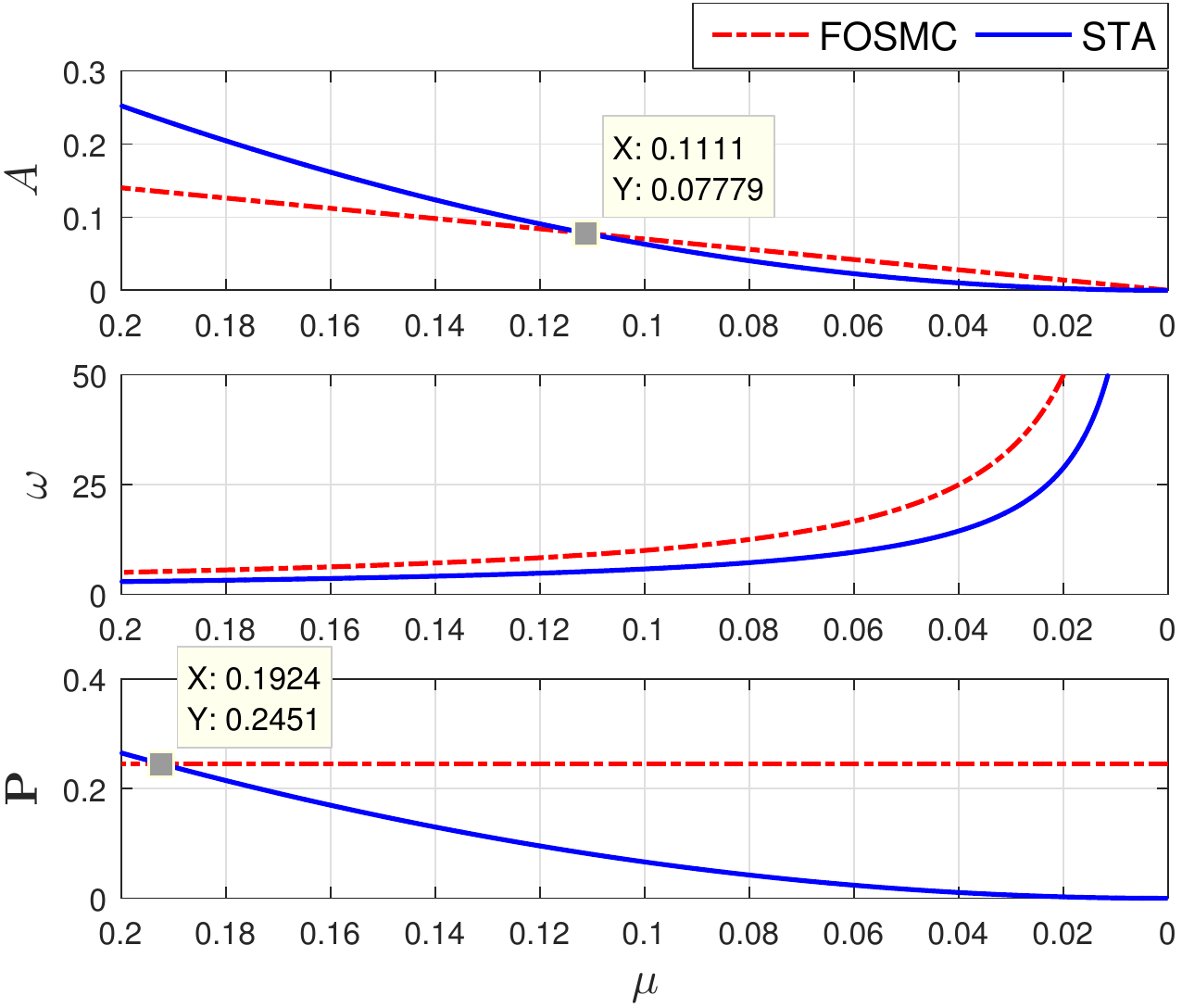}   
\end{center}
\vspace{-5mm}
\caption{Chattering parameters for $\delta = \Delta = 1$ choosing the STA gains (\ref{STA_P_min}).}
\vspace{-4mm}
\end{figure}

\begin{remark}
\textit{Figure 5 confirms that to substitute FOSMC by STA, we should consider that the AP may be greater(lower) when the ATC,}
\begin{equation}
\begin{array}{lcl}
\mu > \mu^\star & \Rightarrow & \text{P}_{\text{FOSMC}} < \text{P}_{\text{STA}} \,, \\
\mu < \mu^\star & \Rightarrow & \text{P}_{\text{FOSMC}} > \text{P}_{\text{STA}} \,. 
\end{array}
\end{equation}
\vspace{1mm}
\end{remark}

\vspace{-4mm}

\section{Comparison Examples}
\subsection{High-Frequency Disturbances}
The previously simulation examples are for the nominal case ($F=0$) but the control gains are selected according to the upperbound of disturbance in the case of FOSMC (\ref{SGN}), or the upperbound of time-derivative disturbance for the STA (\ref{STA}), then
\begin{itemize}
\item \textbf{FOSMC} (\ref{SGN}):
\begin{equation*}
\begin{array}{rclcl}
M & = & 1.1\delta & = & 1.1 \,\alpha \,.
\end{array}
\end{equation*}
\item \textbf{STA} (\ref{STA}):
\begin{equation*}
\begin{array}{rclcl}
k_1 & = & 2.127\sqrt{\Delta} & = & 2.127\sqrt{\alpha \, \Omega} \,, \\
k_2 & = & 1.1\Delta & = & 1.1 \, \alpha\, \Omega \,.
\end{array}
\end{equation*}
\end{itemize}
The value of ATC predicted by HB for which the amplitude of chattering is the same despite the use of discontinuous FOSMC (\ref{SGN}) or continuous STA (\ref{STA}), on the dynamically perturbed system (\ref{System_DinPert}) is
\begin{equation} \label{Critico}
\mu^\ast = 0.125 \dfrac{\delta}{\Delta} = 0.125 \dfrac{1}{\Omega} \,.
\end{equation}
In order to compare the system behavior for some values of the ATC, consider
\begin{equation} \label{Fronteras}
\mu_1 = 0.25 \dfrac{1}{\Omega} \,, \hspace{4mm}
\mu_2 = 0.0833 \dfrac{1}{\Omega} \,.
\end{equation}
Table \ref{Accuracy_min_A} contains the sliding-mode output accuracy for some values of disturbance frequency $\Omega$, taking into account the critical value of ATC (\ref{Critico}) and $\mu_1>\mu^\ast$, $\mu_2<\mu^\ast$ from (\ref{Fronteras}). \\

\begin{remark}
\textit{Simulation results confirm that for any disturbance frequency $\Omega$ should be a critical value of ATC $\mu^\ast$ for which the magnitude of chattering is the same when FOSMC or STA are applied. If ATC is greater than $\mu^\ast$ (for example $\mu_1$) the amplitude of oscillations is lower using FOSMC than the obtained applying STA. But if ATC is lower than $\mu^\ast$ (for example $\mu_2$) the amplitude of oscillations is higher using FOSMC than the obtained applying STA.}
\vspace{1mm}
\end{remark}

\begin{table}[t]
\centering
\scalebox{1.12}{
\begin{tabular}{|c|c||c|c|c|}
\hline
% Signo
\multicolumn{2}{|c||}{\bf \backslashbox{{\tiny Control}}{{\tiny $\Omega$}}} & {\tiny \textbf{1}} & {\tiny \textbf{10}} & {\tiny \textbf{100}} \\ \hline \hline
\multicolumn{5}{|c|}{{\tiny Discontinuous Control}} \\ \hline \hline
% Signo
\multirow{3}{*}{{\tiny \textbf{FOSMC}}} & {\tiny $\mathbf{\mu_{_1}}$} & \begin{tiny} 1.6326 \end{tiny} & \begin{tiny} 1.6224$\times 10^{-1}$  \end{tiny} & \begin{tiny} 1.6226$\times 10^{-2}$  \end{tiny} \\ \hhline{*{1}{~}*{4}{|-}} & {\tiny $\mathbf{\mu^\ast}$} & \begin{tiny} \cellcolor{blue!15} 1.7644$\times 10^{-1}$ \end{tiny} & \begin{tiny} \cellcolor{blue!15} 1.9018$\times 10^{-2}$  \end{tiny} & \begin{tiny} \cellcolor{blue!15} 1.8969$\times 10^{-3}$  \end{tiny} \\ \hhline{*{1}{~}*{4}{|-}} & {\tiny $\mathbf{\mu_{_2}}$} & \begin{tiny} 9.4217$\times 10^{-2}$ \end{tiny} & \begin{tiny} 9.4311$\times 10^{-3}$ \end{tiny} & \begin{tiny} 9.4872$\times 10^{-4}$ \end{tiny} \\ \hline \hline
\multicolumn{5}{|c|}{{\tiny Continuous Control}} \\ \hline \hline
% STA
\multirow{3}{*}{{\tiny \textbf{STA}}} & {\tiny $\mathbf{\mu_{_1}}$} & \begin{tiny} 2.2492 \end{tiny} & \begin{tiny} 2.6933$\times 10^{-1}$ \end{tiny} & \begin{tiny} 2.7061$\times 10^{-2}$ \end{tiny} \\ \hhline{*{1}{~}*{4}{|-}} & {\tiny $\mathbf{\mu^\ast}$} & \begin{tiny} \cellcolor{blue!15} 1.3229$\times 10^{-1}$ \end{tiny} & \begin{tiny} \cellcolor{blue!15} 1.3516$\times 10^{-2}$ \end{tiny} & \begin{tiny} \cellcolor{blue!15} 1.3518$\times 10^{-3}$ \end{tiny} \\ \hhline{*{1}{~}*{4}{|-}} & {\tiny $\mathbf{\mu_{_2}}$} & \begin{tiny} 4.8421$\times 10^{-2}$ \end{tiny} & \begin{tiny} 4.8374$\times 10^{-3}$ \end{tiny} & \begin{tiny} 4.8573$\times 10^{-4}$ \end{tiny} \\ \hline
\end{tabular}}
\vspace{2mm}
\caption{Sliding-Mode Output Accuracy for Minimum Amplitude STA Gains (\ref{STA_A_min}).}\label{Accuracy_min_A}
\vspace{-6mm}
\end{table}

\subsection{Professor V. Utkin Example}
The following example was taken from the paper \cite{Utkin15}, they propose that the upperbound of disturbance and the upperbound of time-derivative disturbance have the same value $\delta = \Delta = 60$. Taking into account the FOSMC (\ref{SGN}) gain $M = 1.1\delta$ and the STA (\ref{STA}) proposed gains (\ref{STA_A_min}), the following chattering parameters are obtained by HB:
\begin{itemize}
\item \textbf{FOSMC}
\begin{equation*}
A = 42.017\mu \,, \hspace{4mm} \omega = \frac{1}{\mu} \,, \hspace{4mm} \text{P} = 882.7102 \,.
\end{equation*}
\item \textbf{STA}
\begin{equation*}
A = 336.135\mu^2 \,, \hspace{4mm} \omega = \frac{1}{\mu\sqrt{2}} \,, \hspace{4mm} \text{P} = 28246.93\mu^2 \,.
\end{equation*}
\end{itemize}

Hence the critical values of ATC become
\begin{equation}\label{Criticos}
\mu^\ast = 0.125 \,, \hspace{4mm} \mu^\star = 0.1768 \,,
\end{equation}
for same amplitude and same AP, respectively. Table V shows the chattering parameters obtained in simulation for some values of ATC and the critical values (\ref{Criticos}). Note that when the ATC $\mu>\mu^\ast$ the amplitude of oscillations generated by FOSMC is lower than the produced by STA, this situation is reversed when $\mu<\mu^\ast$. On the other hand, when the ATC $\mu>\mu^\star$ the AP generated by FOSMC is lower than the produced by STA, and when $\mu<\mu^\star$ the AP caused by FOSMC is grater than the produced by STA.

\begin{table}[t]
\centering
\scalebox{1.12}{
\begin{tabular}{|c|c||c|c|c|c|}
\hline
% Signo
\multicolumn{2}{|c||}{\bf \backslashbox{{\tiny Control}}{{\tiny $\mu$}}} & {\tiny \textbf{0.2}} & {\tiny \textbf{0.1768}} & {\tiny \textbf{0.125}} & {\tiny \textbf{0.1}} \\ \hline \hline
\multicolumn{6}{|c|}{{\tiny Discontinuous Control}} \\ \hline \hline \hhline{{6}{|-}}
% Signo
\multirow{3}{*}{{\tiny \textbf{FOSMC}}} & {\tiny $\mathbf{A}$} & \begin{tiny} 8.6899 \end{tiny} & \begin{tiny} 7.6819  \end{tiny} & \begin{tiny} \cellcolor{blue!10} 5.4312  \end{tiny} & \begin{tiny} 4.3450  \end{tiny} \\ \hhline{*{1}{~}*{5}{|-}} & {\tiny $\mathbf{\omega}$} & \begin{tiny} 4.8900 \end{tiny} & \begin{tiny} 5.5317  \end{tiny} & \begin{tiny} 7.8240  \end{tiny} & \begin{tiny} 9.7800 \end{tiny} \\ \hhline{*{1}{~}*{5}{|-}} & {\tiny $\text{P}$} & \begin{tiny} 926.899 \end{tiny} & \begin{tiny} \cellcolor{blue!20} 926.899 \end{tiny} & \begin{tiny} 926.899 \end{tiny} & \begin{tiny} 926.899 \end{tiny} \\ \hline \hline
\multicolumn{6}{|c|}{{\tiny Continuous Control}} \\ \hline \hline
% STA
\multirow{3}{*}{{\tiny \textbf{STA}}} & {\tiny $\mathbf{A}$} & \begin{tiny} 13.5615 \end{tiny} & \begin{tiny} 10.5999 \end{tiny} & \begin{tiny} \cellcolor{blue!10} 5.2987  \end{tiny} & \begin{tiny} 3.3911 \end{tiny} \\ \hhline{*{1}{~}*{5}{|-}} & {\tiny $\mathbf{\omega}$} & \begin{tiny} 3.5153 \end{tiny} & \begin{tiny} 3.9764 \end{tiny} & \begin{tiny} 5.6242 \end{tiny} & \begin{tiny} 7.0302 \end{tiny} \\ \hhline{*{1}{~}*{5}{|-}} & {\tiny $\text{P}$} & \begin{tiny} 1152.394 \end{tiny} & \begin{tiny} \cellcolor{blue!20} 900.6406 \end{tiny} & \begin{tiny} 450.360 \end{tiny} & \begin{tiny} 288.2422 \end{tiny} \\ \hline
\end{tabular}}
\vspace{2mm}
\caption{Chattering Parameters Obtained by Simulations.}
\vspace{-6mm}
\end{table}

\section{Conclusions}
\noindent
A methodology for analysis based on Harmonic Balance approach is proposed to study the chattering for dynamically perturbed systems driven by FOSMC and STA. HB approach and simulation results allows to confirm the Professor V. Utkin hypothesis, given the value of ATC and the upperbound of disturbance
there exist a bounded disturbance for which the amplitude of possible oscillations produced by FOSMC is lower than the obtained applying STA. On the other hand, given the upperbound of disturbance and the upperbound of time-derivative disturbance there exist an actuator fast enough for which the STA amplitude of oscillations or(and) AP are lower than the generate by FOSMC. Selection criteria of STA gains are presented to adjust the chattering effects on the amplitude of possible oscillations and AP.

\section*{Acknowledgment}
The authors are grateful for the financial support of CONACyT (Consejo Nacional de Ciencia y Tecnolog\'{i}a): CVU 631266; PAPIIT-UNAM (Programa de Apoyo a Proyectos de Investigaci\'{o}n e Innovaci\'{o}n Tecnol\'{o}gica) IN 113612.


\nocite{*}  

\begin{thebibliography}{20}

\bibitem{Atherton75}
D.P. Atherton, \emph{Nonlinear Control Engineering-Describing Function Analysis and Design}, Van Nostrand Reingold: New York and London, 1975.

\bibitem{Bartolini98}
G. Bartolini and A. Ferrara and E. Usani, \emph{Chattering avoidance by second-order sliding mode control}, IEEE transactions on Automatic Control, vol. 43, no. 2, pp. 241--246, IEEE, 1998.

\bibitem{Boiko05}
I. Boiko and L. Fridman, \emph{Analysis of chattering in continuous sliding mode controllers}, Automatic Control, IEEE Transactions on, vol. 50, no. 9, pp. 1442--1446, IEEE, 2005.

\bibitem{Boiko07}
I. Boiko and L. Fridman and A. Pisano and E. Usai, \emph{Analysis of chattering in systems with second-order sliding modes}, Automatic Control, IEEE Transactions on, vol. 52, no. 11, pp. 2085--2102, IEEE, 2007.

\bibitem{Boiko09}
I. Boiko, \emph{Frequency-Domain Analysis and Design}, Birkhäuser Basel, 2009.

\bibitem{Emelyanov86}
S.V. Emelyanov and S.K. Korovin and L.V. Levantovskii, \emph{Higher-order sliding modes in binary control systems}, Soviet Physics Doklady, vol. 31, no. 31, pp. 291,1986.

\bibitem{Fridman01}
L. M. Fridman, \emph{An averaging approach to chattering}, IEEE Transactions on Automatic Control, vol. 46, no. 8, pp. 1260--1265, IEEE, 2001.

\bibitem{Fridman02}
L. M. Fridman, \emph{Singularly perturbed analysis of chattering in relay control systems}, IEEE Transactions on Automatic Control, vol. 47, no. 12, pp. 2079--2084, IEEE, 2002.

\bibitem{Gelb68}
A. Gelb and W. E. Vander Velde, \emph{Multiple-Input Describing Functions and Nonlinear System Design}, McGraw Hill, 1968.

\bibitem{Levant93}
A. Levant, \emph{Sliding order and sliding accuracy in sliding mode control}, International journal of control, vol. 58, no. 6, pp. 1247--1263, Taylor \& Francis, 1993.

\bibitem{Levant98}
A. Levant, \emph{Robust exact differentiation via sliding mode technique}, Automatica, vol. 34, no. 3, pp. 379--384, Elsevier, 1998.

\bibitem{Levant01}
A. Levant and A. Pisano and E. Usai, \emph{Output-feedback control of the contact-force in high-speed-train pantographs}, Decision and Control, 2001. Proceedings of the 40th IEEE Conference on, vol. 2, pp. 1831--1836, IEEE, 2001.

\bibitem{Levant10}
A. Levant, \emph{Chattering analysis}, Automatic Control, IEEE Transactions on, vol. 55, no. 6, pp. 1380--1389, IEEE, 2010.

\bibitem{Moreno09}
J. A. Moreno, \emph{A linear framework for the robust stability analysis of a generalized super-twisting algorithm}, Electrical Engineering, Computing Science and Automatic Control, CCE, 2009 6th International Conference on, pp. 1--6, IEEE, 2009.

\bibitem{Ventura16}
U. Pérez-Ventura and L. Fridman, \emph{Chattering measurement in SMC and HOSMC}, 2016 14th International Workshop on Variable Structure Systems (VSS), pp. 108--113, IEEE, 2016.

\bibitem{Swikir16}
A. Swikir and V. Utkin, \emph{Chattering analysis of conventional and super twisting sliding mode control algorithm}, 2016 14th International Workshop on Variable Structure Systems (VSS), pp. 98--102, IEEE, 2016.

\bibitem{Tsypkin85}
Ya. Z. Tsypkin, \emph{Relay control systems}, Cambridge, U.K.: Cambridge Univ. Press, 1985.

\bibitem{Utkin92}
V. Utkin, \emph{Sliding modes in optimization and control problems}, Springer Verlag, New York, 1992.

\bibitem{Utkin15}
V. Utkin, \emph{Discussion Aspects of High Order Sliding Mode Control}, Automatic Control, IEEE Transactions on, vol. 61, no. 3, pp. 829--833, IEEE, 2016.

\bibitem{Utkin16}
V. Utkin, \emph{Divergence theorem for super twisting control}, 2016 14th International Workshop on Variable Structure Systems (VSS), pp. 178--181, IEEE, 2016.

\end{thebibliography}
\end{document}